\documentstyle[aps,pre]{revtex}

\begin{document}
\title{Spreading, shrinking and fingering instability of polymer films in glow discharge }
\author{Vladimir Kolevzon}
\address{on leave from: Lab de Physique, Ecole Normale Superieure de Lyon\\
46 allee d'Italie, 69364 Lyon CEDEX 07, France}
\maketitle
\begin{abstract}
Spontaneous spreading of a polymer drop on a solid substrate can be forced  by a low pressure
gaseous discharge applied between the cathode that supports the drop and a remote electrode.
We prone to think that the driving force for this plasma driven spreading is the spatial distribution of the electric
field at the cathode. Two distinct effects caused by the gas plasma are observed: spreading
of a thin film and shrinking of the rest liquid into a spherical drop upon its own film.
The last effect, never reported before, can be addressed to the effect of the electric stresses
 developed in the ionic double layer formed at the polymer/plasma interface.
The spreading front displays a fingering instability similar to that observed earlier in
various experimental conditions \cite{troian,melo}. Theoretical analysis shows that the
Marangoni stress acts as the driving force for the finger growth.
\end{abstract}
\section{Introduction}
The spreading of a liquid film on a solid substrate is of great theoretical and practical
importance in such applications like coatings or adhesion. There are several basic
mechanisms providing the driving force for the fabrication of thin liquid films on a
substrate, namely spin up or thermal gradients that drive a thin film along the
 substrate \cite{troian1}.

The present paper reports a novel technique enable to manipulate the spreading of
liquid films at a slightly conductive solid. We reported already that wetting
(spreading) between liquid gallium and a solid substrate at room temperature can be obtained
in a glow discharge \cite{kol}. Liquid layers prepared in such a way remained stable after
switching off the discharge. We concluded that irreversible electric wetting
contrary to the well known electrocapillary effect is observed. We addressed the physical
mechanism of this irreversible spreading to the action of an additional shear force
acting in the external electric field superposed with the glow electric current.
To prevent the liquid metal surface from oxidation that study
was restricted to the glow discharge in a vacuum of about 10$^{-2}$ Torr.

Now we study the spreading of a dielectric liquid on a Si-wafer at
gaseous discharge fired at slightly higher pressure: from  $10^{-1}$ to 1 Torr.
Two novel effects are observed: first, fingering instability of the spreading
front developed at the liquid film/wafer interface. Second, a collapse of the rest of liquid
to nearly perfect sphere upon its own thin film. Both of the two effects are observed
simultaneously and are well reproducible in a series of experiments. Note that the first
effect was reported in a number of experiments: in spin up coating
\cite{melo} or for flows down an inclined plane \cite{troian11}, or in Marangoni-driven
 spreading of surfactants on water \cite{troian}, while
the second effect was never reported (to the author's knowledge).

Theoretical
analysis, carried out in this study, shows that the distribution of the electric stresses
in the ionic layer formed at the liquid/plasma interface is the key point defining the
direction of the net surface force acting on a spherical segment.
Moreover, due to the effect of the strong electric field (focused to the ionic double layer)
making the tension negative (at the liquid/plasma interface)
the breakup of a liquid lens is observed. Clearly, plasma action produces great interfacial
effects on liquid polymers contrary to the reported small changes in the surface properties
of corona treated polymers \cite{bosmina}.  Hence we believe in great technical potential of
 glow discharge plasma helpful for spreading of much thiner films than those
available by any classical technique.

\section{Experimental}
The experimental set up is extremely simple; it consist of a vacuum chamber having
two vacuum flanges of the diameter about 10 cm spaced by the
length of the glass tube about 10 cm. The needle-shaped anode was fixed to the
upper flange and a polished Si wafer was fixed at
the Al sample holder bolted to the lower vacuum flange which served as the cathode.
A small droplet of silicone oil (Rhodorsil 47V500, density $\rho$=0.97 g/cm$^3$, viscosity
$\eta$=500 cSt at 25 C) was deposit on Si wafer,
 rinsed with the surfactants solution and distilled water, prior to use. Liquid
completely wets the substrate and spreads spontaneously in a few minutes to diameter of about
 6 to 10 mm depending on the
initial volume. Polymers of this sort have very low vapor pressure so they are ideal
candidates for vacuum experiments. The particular choice of the liquid viscosity was dictated
by the wish to demonstrate the power of the method for highly viscous liquids; but the
upper limitation is posed by oils of the viscosity $\eta\sim$10$^5$cSt that are almost solid at room
temperature.

The chamber was vacuum sealed and pumped down to
10$^{-2}$ Torr using a vacuum pump (Pfeifer) then pressurized  with N$_2$ through the
leak--valve up to the total pressure about 5$\times 10^{-1}$ Torr. The voltage was applied between
two electrodes and gradually increased up to 1 kV. A glow discharge ignited at this voltage
caused the voltage to fall down to 500 V and the discharge current rose to 10 mA integrally.
The glow discharge appeared in the form of  small luminous space near the needle-shaped anode
(the so-called anode glow \cite{discharg}) and a large dark zone extended to the cathode.
The cathode glow was not always seen at the same place: sometimes it concentrated at
the sample holder,
sometimes the bright streamers were attracted to the rim of vacuum flange. To
concentrate the plasma near the sample holes were drilled in the sample holder that
enhanced locally the electric field and the luminous plasma balls were clearly visible near the
holes.  Note that the present configuration (the needle at '+' and the plane electrode
at '--') was the only one possibility to get the glow above the plane electrode, the maximum
discharge current and the spreading effect. If the electrode polarity is reversed the plasma
glow is concentrated near the needle-shaped cathode, while the anode is usually dark.
It should be stressed that no spreading effect was observed under the reversed polarity.

After the discharge is ignited the drop spreads
in a few seconds with an averaged speed of the front of the order of 1 mm/s.
Such speed is compatible to the front speed achieved in the surfactants spreading on water
governed by the concentration gradient of the surface tension \cite{troian}. In our
case the electric force induced by the electric field gradient at the cathode (see below)
drives the thin film spreading.
The spreading patterns produced in vacuum were recorded by a
CCD-camera with the Olympus macro-objective after the chamber was opened. The images were fed to the computer by
the standard 8-bit frame grabber. A typical spreading pattern is shown in Fig 1. The
initial diameter of liquid prior to the discharge was about 4.5 mm while the final
pattern displayed the fingers with a characteristic length of about 3.5 mm. The finger length
was somewhat dependent on the residual vacuum or plasma conditions: in some cases we
obtained the fingers up to 20 mm long modulated by dendrites grown on them.

In order to demonstrate the importance of the electric field at the cathode the following
experiments are carried out. First, the experimental geometry is changed in such a way that an Al
ring around the sample was bolted to the sample holder. Due to its high conductivity (compared
to Si)
the Al ring attracts the glow current and changes the spatial configuration of the field at
the cathode. As the result of this the spreading patterns have been changed: the finger
length is reduced; sometimes no spreading pattern at all has been visible and only shrinking is
greatly pronounced (see Fig. 2). In a second experiment a special form of glow discharge is ignited at
a hollow anode according to \cite{anders}. The liquid drop is on a flat anode and a
dielectric  disc with a small hole
in the middle is placed a few mm above the anode. The E-field lines are concentrated near the
disc hole that lead to ignition of the bright plasma ball inside the hole. The 'anode' plasma
blown through the anode hole by the pressure gradient forms a plasma jet that should reach the
flat electrode.   In this configuration a dense plasma is focused in the vicinity of the
liquid drop. Surprisingly no changes--either spreading or shrinking are observed  on the
anode. We prone to think that these two results ensure that the electric field at the cathode
serves at the main driving force for the liquid spreading in glow discharge.

In all experiments the spreading of a thin film is accompanied by the
backward motion of the rest of
liquid that forms a drop in the pattern center. This drop, lying
on its own film, had a contact angle nearly 90$^\circ$  that was never reported before (to the
author's knowledge). The drop remained stable in glow discharge and for periods
of time up to some weeks in open air. However it was easy to change the drop contact
angle (only for low viscous liquids) by blowing a gentle stream of air to it.
Significant changes were observed with increasing the liquid viscosity. No spreading
pattern could be obtained for liquid drops of the viscosity $\eta$=2000 cSt; only shrinking
took place despite a very bright plasma balls around the drop was clearly visible.
After the liquid shrinking
small droplets of sub-mm size remained upon a thin film decorating the initial
position of the liquid front. These observations signal the breakup of the liquid layer due to
a negative tension at the liquid/plasma interface  caused by the strong electric field
concentrated in a very thin ionic double layer (see below).


\section{Liquid drop in a glow discharge}
In the theoretical part of this paper we try to find out the basic equations describing the electric forces
acting on the liquid layer in glow discharge. The interfacial electric stresses arise from
the two orthogonal components of the electric field as formulated in the excellent review of
electrohydrodynamics \cite{melcher}. To find out the distribution of the electric field E
 in a most general way the Laplace equation ($\nabla^2\phi=0$) for the electrostatic potential
should be solved inside the liquid layer. This is a non trivial task regarding the
difficulties in the formulation of the boundary conditions at the interface between a slightly
conducting plasma and a highly dielectric liquid. Therefore we will first try to formulate the
boundary conditions at the liquid/plasma interface in order to get possible numeric estimates
of the E-field  and the interfacial stresses consequently.

First, we would like to remind the basic facts pertinent to the glow discharge at
low pressure \cite{discharg}.
It should be stressed that an electric field of any strength applied in a vacuum induces very small current (
typically in the range of 10$^{-10}$ to 10$^{-6}$ A \cite{discharg}) between two electrodes.
However we will neglect this current and suppose to have only a static electric field  prior 
to the discharge. In this situation the field inside the dielectric is $\epsilon$ times
($\epsilon$=2.8--the dielectric permittivity of silicone oil) lower
than E$_z$ in vacuum. The current growth is described by $I=I_0\exp(\alpha d)$, where 
$\alpha$ is the Townsend coefficient and d is the  distance between electrodes. I$_0$ is some 
initial photocurrent caused by n$_0$
electrons emitted from the cathode. These electrons are accelerated by the field and their 
number grows exponentially if the value E/p (p is the chamber pressure) is high enough. In such
 a way an electron avalanche
having the number of electrons $n\propto\exp(\alpha d)$ strikes the anode
and at the same time produces further ionisations due to collisions with gas atoms. Note that
$\alpha/p$ is a strong function of E/p \cite{discharg}. 
At some definite value of the field which is mainly determined by the product 
(pd) the breakdown occurs. The current rises several orders of magnitude to few mA whereas the
voltage drops and the discharge becomes self-sustained.

Various fundamental processes that influence the electric current happen in a glow discharge.
Those are \cite{discharg}: ionisations by electron collisions, the drift motion of electrons 
and ions in the electric field, recombination of ions and electrons and photoionization and 
finally the plasma oscillations making the gas current nonuniform. We will not concentrate on 
every details of 
these contributions to the discharge current but treat the gas plasma as a continuous medium 
having the conductivity $\sigma_p$ much higher than the conductivity of the
liquid $\sigma_p >>\sigma_l$.
In such a way we suggest that the electric current is concentrated around the drop but does not
flow through.

Consider a liquid wedge, whose surface is given by a function h(x), placed on a solid
electrode.
External electric field E$_z$, applied prior to the discharge, is uniform near the electrode and is slightly distorted near the
drop edges. Under the influence of electric field the polarization charges are induced on both
surfaces of the drop whereas the charges in the bulk are absent. So in the case when the
drop is on the cathode a negative charge (polarization charge) is induced on the upper
surface. The lower surface bears a free charge density imposed by the cathode and
polarization charge (positive) equilibrating that charge density on the upper surface.
The field inside the drop is
uniform too and parallel  to E$_z$ and given by $E_d=\delta\phi/h(x),$ where $\delta\phi$ is
the potential difference between the upper and lower surface of the drop.

The boundary conditions for the electric field at the plasma--drop interface are non trivial. 
We suppose that the current perpendicular to the interface $j_n=-\sigma_p\frac{\partial\phi}
{\partial n}$ can be matched to the  drop surface charge density \cite{melcher}
\begin{equation}
\label{field1}
j_n-\sigma_l\frac{\partial\phi}{\partial n}+{\rm div}_s(q_s u_s)=-\frac{\partial q_s}{\partial t}
\end{equation}
where q$_s$ the surface density of the free charges, i.e. those charges that could be transfered
to the bulk by the conduction or convection current originating from the surface velocity
u$_s$.
This boundary condition  can serve for a very rough estimate on the surface velocity if both
the time dependent term and the current inside the drop are neglected: $u_s\sim j_n/q_s/h$ where h is taken as a
characteristic length scale for the changes in surface velocity. The order of magnitude
estimates of the plasma current density on the drop surface $j_n \sim 10^{-3}$ mA/cm$^2$
and the surface free charge density (created by the plasma ions) $q_s \sim 10^{-6}$ Q/cm$^2$
give the surface velocity $u_s\sim j_n h/q_s=10^{-6} \times 10^{-1}/10^{-6}$=1 mm/s.
Of course, for more exact estimates one needs to solve the Laplace equation in both media
with the boundary condition in the form of Eq(\theequation).
The time dependent term on the RHS describes the rate of production of the electric
charges at the liquid surface; usually the change in the volume charge density q (ions or electrons)
is given by $\frac{dq}{dt}\propto R,$ where R is the pressure dependent dimensional
coefficient $\propto 1/p$.
R corresponds to the particular process in glow discharge taking place near the drop surface:
for instance the electron-ion recombination or electron collisions (R=[qP(u)]-the ionisation
probability) or drift motion of ions (R=q div W, W is the drift velocity).
The component of the electric current inside
the liquid $\sigma_l \frac{\partial\phi}{\partial n}$ is greatly influenced by the liquid velocity
u$_s$.

This point needs  further comments. The ratio of both
contributions (convective/conductive) is given by \cite{melcher}
$$uq/l\sigma\nabla\phi=\frac{\epsilon_0 V_0/l u}{l\sigma V_0/l}=
\frac{\epsilon_0 u}{\sigma l}=R_e$$ and is
called the electric Reynolds number R$_e$. Using typical conductivity of silicone oil 
$\sigma=10^{-14}$ (Om m)$^{-1}$ and characteristic length of 1 mm (a thickness of liquid
layer) one gets
$R_e=10^{-11}\times 10^{-3}/(10^{-14}\times 10^{-3})=10^3$ even for the surface
velocity u$_s \sim$1 mm/s.
Hence the normal component of the field inside the drop E$_n$ is influenced by the fluid
motion but the generation of the surface charges seems to be more quick being a very
important parameter affecting the field E$_n$. Another
notion is that the decay time of the polarization  charges is rather high: for silicone oil of the
conductivity the surface charges decay to their 50\%
magnitude for $\tau=\epsilon_0/\sigma$=1000 s. Hence we can conclude that the {\bf E}-field pattern inside the drop remains
unchanged for some time.

According to the formulation of Ref \cite{melcher} the electric interfacial stress is given
by the product of the two orthogonal field components
\begin{equation}
S_{el}=\epsilon_0 E_\perp E_\parallel
\end{equation}
 where the E-component parallel to the surface
has to be deduced from the other boundary condition. It is easy to imagine the situation when the glow current
streams around the curved liquid layer like the hydrodynamic flow around a sphere.
The continuous changes of the tangential electric field $E_\parallel$ are expressed as
\begin{equation}
E_\parallel-j_\parallel/\sigma_p=0
\end{equation}
 In such a way the tangential component
of the field (absent prior to discharge) is developed on the drop surface due to the 
distortion of the field pattern by the glow current (see Fig 3). This tangential component is zero at the
 lower surface where only the normal component $E_n$ is proportional to the free (f) and
polarization (p) charge density $\epsilon \epsilon_0 E_n=(Q_f-Q_p).$ It
 is not easy even to give the order of magnitude estimates of $j_\parallel$
 because only a small fraction of the total current flows parallel to the drop surface and
 the estimations of $E_\parallel$ fail.
\section{Spreading versus shrinking}
In this section we will collect some theoretical facts able to explain spreading and
shrinking of the liquid layer into a spherical drop. Although these two effect are observed
simultaneously the physical mechanisms are very different and reflect the
basic processes in the glow discharge. As shown in the previous section the standard
approach to the interfacial electric stresses failed to give the estimates on the surface
force acting at the plasma/liquid interface.

An alternative way how to calculate the spreading force is to account for
the main feature of the glow discharge namely the cathode dark space. This is a region 
having the potential drop across it of the order of several hundreds volts. Large 
electric field concentrated at the length of some mm is the key element for driving force
acting on the liquid drop. Indeed the field E$_l$ inside the drop is preserved even after 
the discharge starts (see above) and is $\epsilon$-times lower than the field in vacuum.
Therefore the electric force perpendicular to the drop edge \cite{landau}
\begin{equation}
\label{spread}
f_{el}=\epsilon_0(\epsilon-1)\nabla_x E_z^2
\end{equation}
(x-axis is directed along the electrode) acts outwards where the electric field is the
highest (see the sketch in Fig. 3).
The surface stress is given by the integration of the volume force over the liquid height
$\Pi_{el}=\int dz f_{el}.$
In such a way we consider the field inside the drop essentially the same as that 
prior to the discharge but the field outside is simply the field in the cathode dark space
 which surrounds the drop on the cathode.
We assume for simplicity the uniform field that gives about 50/$\epsilon$ V/cm inside the
drop (the maximum field established prior to discharge) whereas the field 100 V/cm in the cathode dark space. Then the electric stresses estimated
for the typical layer thickness of 1 mm
$\Pi_{el}\approx\epsilon_0(\epsilon-1)\delta E^2 h/\delta L\sim 10^{-11}\times(10^4)^2=$1 mN/m$^2$, where
$\delta L$--the characteristic length of the changes in E is $\delta L\sim h$. The
effect of this electric force can be compared with the Marangoni stresses developed, for
instance, by the temperature derivative of the surface tension $\mid\partial\gamma/
\partial T\mid$=0.1 mN/m/K (\cite{adam}) in the thermal gradient of 1 K/cm
$\nabla\gamma=
\frac{\partial\gamma}{\partial T}\frac{\partial T}{\partial L}=0.1/10^{-2}$=10 mN/m$^2$.

As we mentioned above the upper drop surface is negatively charged. These are pure 
polarization charges; they are closely linked with the electrode potential. These
surface charges attract positive ions and electrons from plasma in such a way that a double
 layer is formed near the liquid surface. The problem of a solid particle
attracting the counter ions from  electrolyte solution has been extensively treated in the 
bulk of literature (see for example \cite{adam}); behavior of a {\it liquid} drop bearing the ionic double layer is much
less known. 

Levich treated this problem in his book. He mentioned two surface forces arising from
the ponderomotive
electric stresses that act in the double layer; we will skip over the details here and 
give only the final equations describing electric forces. The first is the tangential 
component acting in the spherical layer given by \cite{levi}: $$ S_t=\frac{1}{R_0}C\delta\phi
\frac{\partial\phi}{\partial\theta}=\frac{1}{R_0}\frac{\partial\gamma}{\partial\theta}
,$$ where $\theta$ is a polar angle in the spherical coordinate system,
C is the capacitance of the double layer per unit area, $\delta
\phi$ is the potential drop there and R$_0$ is the drop radius. This component $S_t$ is 
fully equivalent to the Marangoni stresses $S_m=\nabla\gamma=S_t$ \cite{levi}.
In addition the normal stress component gives the reduction in the
surface energy $S_n=\delta\gamma=-\frac{C\delta\phi^2}{R_0}.$
Note that both S$_t$ and S$_n$ are $\theta$-dependent because the electric potential of
the double layer $\delta\phi$ is greatly influenced by $\theta$. If the stress
distribution is integrated over the whole spherical surface the resulting force is zero 
\cite{levi}: $\int_0^{\pi} d\theta(S_t-S_n)=0$. 

However this is not so when a spherical segment only is considered. The resulting surface 
force acting in the horizontal direction (x) on a half-spherical segment enclosed
between the angles from 0 to $\theta$
is given by  the integration of the normal and tangential surface stresses \cite{levi}
$f_x=\int_0^{\theta} d\theta \sin\theta(S_t\sin\theta-S_n \cos\theta)$,
where the spherical coordinate system is built upon the radius of curvature of the segment.
In order to escape various complications connected with detailed distribution
of electric field we express the force via the reduction in
the surface tension $\delta\gamma$ \cite{levi}:
\begin{equation}
f_x=\int_0^{\theta} d\theta\sin\theta(\frac{\partial\delta\gamma}{\partial\theta}
\sin\theta+\delta\gamma\cos\theta)=\delta\gamma\sin^2\theta.
\end{equation}
It is evident that the surface forces applied to the two half-spherical segments
$[-\theta;0]$ and $[0;\theta]$ are equal in the magnitude but opposite in sign. These two forces f$_x$ and
-f$_x$ produce the shear stresses  applied to the segment in the horizontal direction.

As we already know the reduction in the surface tension is proportional to the electrostatic
 energy of the double layer therefore the tangential force at the liquid surface can be 
written as $ f_t \sim C(\delta\phi)^2\sin^2\theta\cos\theta,$ where the capacitance of the
double layer and its surface charge density are linked as $Q_s=C\delta\phi$.
There seems to exist a non-electric surface force whose impact would be to compensate the
electric stresses S$_n$ and S$_t$. This surface stress usually proportional to the squared
density gradient $\propto
(\frac{d\rho}{d z})^2$ acts, for example, at the free liquid metal surfaces and stems from
the anisotropy of the density profile in the surface zone \cite{kol1}. In such a way
all surface stresses are balanced at a free
liquid surface in equilibrium with its intrinsic double layer, which is not the case in our
study. Due to a constant flux of ions delivered from the gas plasma no
equilibrium exists at the interface and the electric stresses prevail over the force
induced by the density gradients.

In the case of a liquid wedge confined by the contact angle $\theta$ and having
the curvature radius R the horizontal force
acting at the ionic double layer can be expressed as
\begin{equation}
f_x=\epsilon_0 E_z E_x=Q_s\frac{\partial\phi}{\partial x}\cos\theta\approx Q_s \phi\frac{\sin\theta}
{R\cos\theta}\approx C(\delta\phi)^2\nabla^3 h
\end{equation}
where the potential distribution inside the wedge is taken in the form
$\phi=E_0 R\cos\theta$ and the polar angle is presumed to be equal to the contact angle
($\theta\approx\theta_c$); also the approximations $\cos\theta \sim 1$ and $\frac{1}{R}\sim
\nabla^2 h$ were used.
The derivative $\frac{\partial\phi}{\partial x}\neq 0$ only at the drop edge; at the flat part the tangential component is absent and only
the normal stresses remain.  In this case the surface tension at the interface liquid-vacuum
becomes non-uniform. The tangential surface force acts along the curved interface in the direction
where the tension is the highest, or the reduction $\delta\gamma=
\int_{-\infty}^{\infty}dz (S_n-S_t)< 0$ (due to the electric
stresses) is the lowest and causes the liquid layer to collapse into a spherical drop.

Significant changes are observed at liquid layers of the viscosity $\eta$=
2000 cSt: the spreading pattern is hardly visible and only shrinking is detected.
It becomes evident that the 'shrinking force' takes over the spreading force for
4 times more viscous liquid layers. There is no simple explanation of this fact;
the influence of the surface curvature is excluded because the surface tension
remains almost the same (20 to 21 dyn/cm) for silicone oils of the two types. The electric
field outside the drop cannot be changed by the liquid viscosity then only the field inside
the liquid layer is affected  by the motion which is viscosity dependent. Consequently
the liquid motion has significant effect on the electric force as expected in the case of
the electric Reynolds number $R_e >>1$. Unfortunately it is very difficult to predict this
influence in the framework of our qualitative analysis.

The tangential force at the charged interface is not the single effect produced by the
surface ionic layer. Even a stronger effect is a significant reduction
in $\gamma$ (with the possible negativity)
due to the strong electric field in the ionic layer. It should be emphasized that negative
$\gamma$ promotes the growth of the interfacial disturbances in the absence of any
retarding force except of the van der Waals attraction to the substrate.
As shown long time ago \cite{teilor} the liquid drop becomes unstable in the strong electric field
where the criterion of instability is $\gamma /R=\epsilon_0 E_n^2$, if only the normal
field component is considered.
The liquid instability appeared in the classical experiments
in the form of a thin jet or spikes developed at the liquid drop surface in the external field
of $ER^{1/2}\approx 4\times 10^3$ V cm$^{-1/2}$ but no breakup was reported \cite{teilor}.
The effect of the breakup of the liquid into small sub-mm sized droplets can be explained
due to the destabilization of a liquid lens in the strong E-field (about 10$^5$ V/cm
in the double layer assuming the potential drop there of the order of 100 mV and the thickness
of 100 A). This electric field is few orders of magnitude
 above the critical one which is necessary for the classical instability onset.
\section{Fingering instability in fast spreading}
In this section we concentrate only on the spreading and a semi-quantitative analysis of
the fingering instability will be given.
First we will give a theoretical form describing the propagation speed of the moving
liquid front in the course of spreading.
As we discussed already the present effect of forced spreading is related to the Marangoni
stresses induced in a glow discharge. Then our situation strongly resembles the case of 
spreading of surfactants on water surface discussed in \cite{troian}. The authors
of Ref \cite{troian} propose a
theoretical model describing the speed of the propagation of the surfactants front. 
Unfortunately
there are several basic assumptions about the specific structure of the height profile 
that restrict the usage of the analytical model proposed in \cite{troian}. Therefore we will
use the results of theoretical predictions that are applied in more general case and are 
summarized in \cite{oron}. The authors consider the situation when the external stresses 
$\tau$ are applied tangentially to the liquid-gas interface. The basic equations governing 
the evolution of thin films are written in the so-called lubrication approximation. Following \cite
{oron} we remind the basic assumptions of this theoretical formalism. The first is that the 
wavelength of perturbations is much large than the film thickness so that the parameter
 $\kappa=h/\lambda\ll 1$. Then the hydrodynamic variables: velocity and pressure are written in power series
of the small parameter $\kappa$ and the terms proportional to non zero powers of
$\kappa$ are omitted.
It becomes apparent that the nonlinear terms can be omitted because they are an order of 
magnitude smaller than the dominant viscous terms \cite{oron}.

This approach applied to the steady flow driven by the shear surface stress $\tau$ gives
the basic equation describing the
speed $U=u_x=\partial_t h$ of the spreading front of a thin film whose thickness depends on
the distance h=h(x) \cite{oron}
\begin{equation}
\eta\partial_t h=\eta U=-\tau\partial_x h -\frac{\gamma}{3}\partial_x(h^3 \partial_x^3 h),
\end{equation}
where $\eta$ is the liquid viscosity. Negative sign on the RHS of Eq(\theequation)
reflects the fact that the spreading velocity is directed towards a decreasing contact angle
which is very small but non zero: $\theta_c=\partial_x h\neq 0.$ The first
term in Eq(\theequation) is proportional to the film thickness indicating $u\propto h$.
$\gamma$ is considered as the tension of upper film on a {\it liquid} substrate in
\cite{troian}
but a different form $\gamma_0 \cos\theta_c^{3-m}$ is suggested for the
description of the same term involved in the film spreading on a {\it solid} in \cite{oron}.
Hence we keep $\gamma=-S+\frac{1}{2}\gamma_0\theta_c^2$ \cite{degen} as
the 'effective' tension of the liquid film, that comprises two terms: one
is the tension of the liquid drop $\gamma_0$ acting inwards, the second is the
spreading coefficient S reflecting the balance between three tensions near
the contact line \cite{degen} that governs the spreading without any extra force.
The magnitude of $\tau$ is equal in our case to the electric stress $\tau=\Pi_{el}
=\epsilon_0(\epsilon-1)\int dz\nabla_x (E_z^2)$.
Note that the propagation velocity U expressed by Eq(\theequation) is given by a very similar
equation (up to some numerical coefficients) proposed in \cite{troian} for the spreading of surfactants on water driven by the
Marangoni effect expressed in terms of the concentration gradient of the tension.
Numerical integration of Eq(\theequation) can give time development of the height profile
and the speed of the propagating front. However this task is far beyond our approximate
study and we will skip directly to the fingering instability analysis.

Unfortunately the one dimensional Eq(\theequation) is not suitable for the 2--D instability analysis.
We will follow the most comprehensive formulation of the problem of stability of thin 
liquid films under different forces given in 
\cite{oron}. Hydrodynamic equations describing the evolution of the 2--D (in the 
surface plane) disturbances at a thin film subject to the shear surface stress 
$\tau$ and surface 
tension can be written in a general form as \cite{oron}:
\begin{eqnarray}
\label{2D}
\eta\partial_t h+\nabla(\tau\frac{h^2}{2})+\gamma\nabla\frac{h^3}{3}
\nabla^3 h=0,
\end{eqnarray}
where  ${\bf\nabla}=(\frac{\partial}{\partial x},\frac{\partial}{\partial y}).$
To make the stability analysis it would be highly desirable to write $\tau$ in the form
of the surface tension gradient:
$\tau=\nabla_x\gamma=\frac{\partial\gamma}{\partial h}\frac{\partial h}{\partial x}\approx 
\delta\gamma\nabla h,$ where $\delta\gamma$ is the change in the surface tension over the
typical distance of $k^{-1}$. To find $\delta\gamma$ we integrate the surface electric stress
$\Pi_{el}$ over the tangential coordinate:
\begin{equation}
\delta\gamma=\epsilon_0(\epsilon-1)\int dx
h(x)\nabla_x (E_z^2)\approx\epsilon_0(\epsilon-1)h(x)\delta(E_z)^2
\end{equation}
where $\delta(E_z)^2$ is the
difference between the electric fields outside and inside the drop.
Note that $\delta\gamma >0$ because the electric field is always higher outside the liquid.
  
The main feature of fingering 
is that the wavevector of surface modes is developed transverse to the propagating front. 
We consider disturbances that create sinusoidal like corrugations of the liquid height 
in the transverse direction and at the same time these oscillations are spatially 
decaying along {\bf u}--the propagation velocity. For simplicity we consider a plane propagating
front whose height
is dependent on two coordinates h=h(x,y). The fluid thickness h comprises two parts: 
one steady h$_0$ 
and one oscillating in time and space as $h'=h_0' \exp(ik_y y+\beta x+st)$ where s is the 
temporal growth rate, k is the oscillation wavenumber along the propagating front and 
$\beta < 0$ is the spatial decay along {\bf u}. As postulated in \cite{troian11,brener} the finger height changes
non monotonically along x with the maximum near the finger tip. Hence the perturbations of
the liquid height should incorporate the spatial wavenumber k$_x$. However due to the strong
damping of these modulations ($\beta>>k_x$) k$_x$ can be excluded from our
analysis. Eq(\ref{2D}) re-written in projections yields:
\begin{equation}
\eta\partial_t h=\eta u_x\partial_x h-\nabla_y(\delta\gamma\nabla_y \frac{h^2}{2}+
\gamma h^3\nabla_y\nabla_y^2 h)-\gamma\nabla_x h^3\nabla_x\nabla_x^2 h
\end{equation}
where the relations $\partial_t h=\partial_t h_0+\partial_t h'$
and  $\partial_t h_0=-{\rm div}(uh)=-u_x \partial_x h$ (u$_x$ is the front speed)
are used and $\delta\gamma$ was introduced above.

Following the analysis of the instability of moving contact lines \cite{brener} we interpret
the terms in Eq(\theequation): $u_x \partial_x h$--is the convective term which is proportional
to the squared height of the contact line (u$\propto$h). The other 
terms in the RHS are the diffusive terms driving the liquid transverse to the main flow
\cite{brener}.
This can be understood from the consideration of the flow between finger tip and 
troughs. We assume that the transverse flux decreases
the liquid height at the troughs then the tip height is larger than the trough
height (see FIG. 4 for the graphic interpretation). The liquid is driven
to the tip by stress component perpendicular to u$_x$ expressed by the term
$\delta\gamma\nabla_y h^2/2$ and directed along the increasing height. Since the velocity
$\propto$h the tip moves faster than the troughs that results in the finger growth.
We see that in our case the main
destabilizing effect comes from the Marangoni stresses.
The surface tension drives the liquid in the transverse direction inwards the finger; 
its contribution 
is a 'diffusive type' with the diffusion coefficient $\gamma h^3$: in the limit of 
$h\rightarrow 0$ it becomes small. 
Unfortunately it is not possible to deduce the most unstable wavenumber k because
there are two unknown k and $\beta$ but only the single equation Eq(\theequation).

Hence we will take
a semiquantitative analysis following the original approach to the finger growth
 developed in \cite{brener}. As emphasized in Ref \cite{brener} the growth of the finger
volume $M=hL(t)\lambda/2$ in
early time is supposed to follow the equation: $ \dot{M}\sim\Phi_y L(t)$
where $\Phi_y$ is the y-component of diffusive and convective flux and L(t) is the finger
length.  Later the finger growth is damped when L(t) reaches some 
characteristic length. Since $M(t)=h\lambda L(t)$ and thus $s=\Phi_y/(h\lambda)\cong
\nabla_y\Phi_y/h$, where
s is the growth rate of the perturbations: ($M=M_0\exp(st)$). From this analysis the growth 
rate of the finger is driven by the y component of the flux (whose convective part is
given by $\Phi_{y(c)}=-h\nabla_y\cdot[u_y h]$:
\begin{equation}
\Phi_y= -h_0 u_y \nabla_y h-\frac{1}{\eta}(\delta\gamma\nabla_y \frac{h^2}{2}+\frac{1}{3}
\gamma h^3\nabla_y\nabla_y^2 h)
\end{equation}
In order to find the transverse velocity component we use the equation of the mass balance
$u_x\mid\beta\mid \lambda/2\simeq u_y $ where $\beta$ is the spatial wave damping in the x
direction. As we mentioned above the two quantities $\beta$ and $k_y$ cannot be found 
from a single equation describing the evolution of perturbations; only the 2--D equation
for the spatial modulations can deliver the information about both  $\beta$ and $k_y$. However
such kind of solution is far complicated to be done in the present analysis; therefore we
use a simple assumption $\mid\beta\mid\approx k_y$ which might overestimate the magnitude
of $u_y$ and the growth rate. Finally the temporal growth rate is
\begin{equation}
s(\lambda)\simeq \frac{u_x h_0}{\lambda^2}+\frac{1}{\eta}(\frac{\delta\gamma h_0}{\lambda^2}-
\frac{h_0^3\gamma}{3\lambda^4})
\end{equation}
The most unstable mode $\lambda_0$ can be obtained by maximizing $s(\lambda)$:
\begin{equation}
\lambda_0^2= \frac{\gamma h_0^3}{3\eta}\frac{1}{(u_x h_0)/2+h_0\delta\gamma/(2\eta)}\sim
\frac{h_0^2}{Ca+Ma_*}
\end{equation}
where Ca=$\eta u/\gamma$ is the capillary number and Ma$_*$ is the 'modified
Marangoni' number given by the normalization $\frac{\delta\gamma}{\gamma}$.
Unfortunately it is difficult to compare theoretically estimated wavelength with the
experimental one because the lack of data on $\delta\gamma$ and unknown h$_0$ to the present
preliminary study.

In sum, we studied the spreading of a polymer film on a conductive solid substrate under
the effect of a glow discharge. The liquid drop is driven by the electric stresses developed
in the glow discharge at the polymer/plasma interface. Several novelties are observed:
first, spreading of a thin liquid film accompanied by fingering instability; second,
shrinking of the rest of liquid into a spherical drop upon its own film and third, the
breakup of the liquid layer into small sub-mm drops on top of a thin wetting film. We try
to explore all these effects  on basis of the interfacial electric stresses induced in
glow discharge. It seems to be that the gradient of the electric field in
the cathode dark space $\nabla_x E_z^2$ serves as the main driving force for the drop
spreading. The spreading front undergoes fingering instability due to the Marangoni stress
driving the liquid perpendicular to the propagation speed. Theoretical analysis of the finger
growth shows that the exact estimates on the most unstable wavelength can be obtained only
from the solution of 2--D equation for the spatial disturbances. However an approximate
analysis shows that $\lambda_0$ strongly depends on the main driving forces
$\delta\gamma$ and U and the physical properties of liquid film like viscosity and the film thickness.
The shrinking effect can be explained due to the electric stresses in the interfacial
ionic double layer. These stresses produce a net horizontal force proportional to the
reduction in the surface tension $\delta\gamma$ and the curvature: $f_x \sim
\delta\gamma\nabla^3 h.$ This horizontal force tends to transform the liquid segment into a
sphere that experiences zero net force. The breakup of the liquid layer into small droplets
is much less known and is related to the negativity of the
surface tension in strong electric field developed in the surface ionic dipole layer.
In the author's opinion all the reported effects deserve further theoretical and
experimental studies.
\section*{Acknowledgments} The author is grateful to the CNRS for the financial support.
The hospitality of the department of Physics at the Ecole Normale Superieure Lyon
where this work was done is gratefully acknowledged. Fruitful discussions were shared
with J F Palierne.

\section{Captions}
Fig.1  Drop spreading and fingering pattern of silicone oil on a Si-wafer
obtained in glow discharge fired  between the wafer (cathode) and a remote anode. The outer
diameter of the pattern is about 12 mm while the inner diameter of the drop prior to
discharge is about 4.5 mm. The liquid shrinking happened in the form of the drop in the
middle is not clear visible.\\

Fig 2. The spreading pattern after the action of a glow discharge on a sample surrounded
by an Al ring (see the text). The finger length is noticeably reduced compared to Fig. 1,
the outer diameter of the patter is approximately 8 mm. Two effects:
spreading and shrinking of a liquid layer into a spherical drop in the center are clearly
visible. Small droplets decorating the initial position of the liquid front appeared as a
result of the breakup of the liquid layer due to the negative tension at the liquid/plasma
interface.\\

Fig 3. Experimental geometry showing the effect of the electric force developed in the
cathode dark space. The liquid segment whose curvature is exaggerated placed on the flat
negatively charged cathode while the remote anode is shown out of the scale. The free
(in circles) and
the polarization charges (without circles) concentrated near the both interfaces are also shown.
The electric force is determined by the field gradient along the drop
edge and is directed outwards where the field E$_0$ is higher than the field inside
the liquid E$_l$. The field distribution near the electrode is supposed to be uniform while
the field pattern inside the drop disturbed by the glow current j is non-uniform.
The positive ions delivered by the glow current, shown at the upper
liquid surface, together with the polarization charges form a double layer where
the  electric stresses are developed.\\

Fig 4. A sketch of the finger growth. The finger profile changing non monotonically along
u$_x$ is shown with the bold line. The
liquid film is driven from the trough to the tip by the electric stress $\tau$ having large
y-component. The surface tension $\gamma$ is directed inwards the finger. Both effects
contribute to the transverse flux $\Phi_y$ that establishes the finger time evolution.\\

\end{document}